\PassOptionsToPackage{square,comma,numbers,sort&compress}{natbib}
\documentclass[final,3p,times,twocolumn]{elsarticle}

\usepackage{amssymb}
\usepackage{amsmath}

\usepackage{hyperref}
\usepackage{graphicx}
\usepackage{color}
\usepackage{bm}
\usepackage{gensymb}
\usepackage{bigints}
\usepackage{pifont}
\usepackage{aas_macros}
\usepackage{physics}
\usepackage{longtable}
\usepackage{threeparttable}

\setcitestyle{square, comma, numbers,sort&compress, super}

\hypersetup{
    colorlinks=true,
    linkcolor=Blue,
    filecolor=Blue,
    urlcolor=Blue,
    citecolor=Blue,
   pdftitle={Hubble constant}
}

\journal{Physics of the Dark Universe}

\begin{document}

\begin{frontmatter}



\title{Fast Radio Bursts as probes of the late-time universe: a new insight on the Hubble tension} 


\author[label1,label2]{Surajit Kalita} 
\ead{skalita@astrouw.edu.pl}
\author[label2]{Shruti Bhatporia} 
\ead{shrutibhatporia@gmail.com}
\author[label2,label3]{Amanda Weltman} 
\ead{amanda.weltman@uct.ac.za}

\affiliation[label1]{organization={Astronomical Observatory, University of Warsaw},
            addressline={Al. Ujazdowskie 4},
            city={Warsaw},
            postcode={00478}, 
            country={Poland}}
\affiliation[label2]{organization={High Energy Physics, Cosmology and Astrophysics Theory (HEPCAT) Group, Department of Mathematics and Applied Mathematics, University of Cape Town},
            addressline={Rondebosch},
            city={Cape Town},
            postcode={7700}, 
            country={South Africa}}
\affiliation[label3]{organization={African Institute for Mathematical Sciences},
            addressline={6 Melrose Road, Muizenberg},
            city={Cape Town},
            postcode={7945}, 
            country={South Africa}}

\begin{abstract}
Fast Radio Bursts (FRBs) are bright radio transient events, a subset of which have been localized to their host galaxies. Their high dispersion measures offer valuable insights into the ionized plasma along their line of sight, enabling them to serve as probes of cosmological parameters. One of the major challenges in contemporary cosmology is the Hubble tension -- an unresolved discrepancy between two independent methods of determining the Universe's expansion rate, yielding differing values for the Hubble constant. In this study, we analyze a sample of 64 extragalactic, localized FRBs observed by various telescopes, employing Bayesian analysis with distinct likelihood functions. Our findings suggest that FRBs serve as tracers of the Hubble constant in the late-time Universe. Notably, our results exhibit smaller error bars compared to previous studies, and the derived Hubble constant with 1$\sigma$ error bars no longer overlap with those obtained from early-Universe measurements. These results underscore the continuing tension between early- and late-time measurements of the Hubble constant.
\end{abstract}




\end{frontmatter}



\section{Introduction}

Fast Radio Bursts (FRBs) are short-lived yet exceptionally energetic radio transients originating from mostly extragalactic sources. First reported in 2007 \cite{2007Sci...318..777L}, FRBs remain an enigmatic phenomenon in astrophysics. These bursts typically persist for only milliseconds, but release immense energy, exhibiting peak flux densities on the order of several Janskys. To date, FRBs have been observed over a radio frequency range of roughly 100\,MHz to 8\,GHz \cite{2021ApJ...911L...3P,2018ApJ...863....2G}. High dispersion measures (DMs) associated with FRBs, a measure of the electron density along the line of sight, offer insights into the intervening medium in the line of sight and thus their distances. These DMs strongly suggest an extragalactic origin for most FRBs. A singular, notable exception exists, namely FRB\,20200428A, which is confirmed to have originated from the Galactic magnetar SGR\,1935+2154~\citep{2020PASP..132c4202B,2020Natur.587...59B,2020Natur.587...54C}. Although the intervening interstellar and intergalactic medium (IGM), in general, introduces uncertainties in these distance estimates, several FRBs have been localized to host galaxies. For instance, the repeating FRB\,20121102A was traced to a dwarf galaxy at a luminosity distance of $972\rm\,Mpc$ \cite{2017ApJ...834L...7T}.

Detections of over 800 FRBs have been reported to date, with the majority identified through the Canadian Hydrogen Intensity Mapping Experiment (CHIME) radio telescope\footnote{\url{https://chime-experiment.ca/en}}. However, there is a growing number of events from other radio observatories, such as the Australian Square Kilometre Array Pathfinder (ASKAP)\footnote{\url{https://www.atnf.csiro.au/projects/askap/index.html}} and MeerKAT\footnote{\url{https://www.sarao.ac.za/science/meerkat/}}, reflecting a rapid increase in discovery rates. Most FRBs are detected as singular, non-repeating events, with no apparent repetition from the same source. Note that many FRBs might have been missed during detection due to detector sensitivity with respect to minimum flux and luminosity of the event. Nevertheless, a small subset of FRBs exhibit repeating behavior, with repeated bursts originating from a consistent location, facilitating in-depth source characterization. FRB\,20121102A is a well-known example of a repeating FRB, with over hundred bursts observed to date. The variability in repetition rates across observed FRBs suggests multiple potential astrophysical origins or mechanisms, posing challenges for a unified theoretical framework \cite{2019PhR...821....1P}. Studying FRBs provides crucial insights into high-energy astrophysical processes and the extreme environments surrounding compact objects such as neutron stars and black holes \cite{2023MNRAS.520.3742K}.

Despite significant progress, the exact origins of FRBs remain uncertain, although several hypotheses have been advanced. The hypothesis of a magnetar origin has garnered substantial support \cite{2014MNRAS.442L...9L}, yet unresolved questions remain, as other mechanisms can also explain many observed FRBs. Notably, only one FRB has been definitively linked to a Galactic magnetar. Consequently, alternative theories are being considered, with some models proposing connections to black holes or white dwarfs. For comprehensive reviews on potential progenitor mechanisms, see \cite{2019PhR...821....1P} and \cite{2020Natur.587...45Z}. Future multi-messenger observations, incorporating gravitational wave (GW) and neutrino detections associated with FRB sources, offer promising avenues for uncovering the astrophysical processes responsible for FRBs \cite{2023MNRAS.520.3742K}.

FRBs, traveling across significant cosmological distances, provide a valuable tool for probing the IGM, including the distribution of matter within it \cite{2019ARA&A..57..417C}. Their unique characteristics, such as short pulse durations, high DMs, and the ability to probe the IGM on cosmological scales, make them effective in addressing various astrophysical and cosmological questions. For instance, FRB\,20150418A has been used to place constraints on the photon mass, limiting it to $m_\gamma<1.8\times 10^{-14}\rm\,eV\,c^{-2}$ \cite{2016PhLB..757..548B}. Subsequent studies of multiple FRBs have strengthened this constraint over the years \cite{2021PhLB..82036596W,2023MNRAS.520.1324L}. In addition, the application of gravitational lensing to FRBs has allowed for constraints on the fraction of dark matter that may consist of primordial black holes \cite{2016PhRvL.117i1301M,2020ApJ...900..122S,2020PhRvD.102b3016L,2020ApJ...896L..11L,2022PhRvD.106d3017L,2023JCAP...11..059K}. Moreover, by marginalizing over the contributions from both the host galaxy and free electrons in 12 FRBs, constraints were put on the parameterized post-Newtonian parameter to test the weak equivalence principle and found that it is consistent with unity within 1 part in $10^{13}$ at the 68\% confidence level \cite{2023MNRAS.523.6264R}, representing one of the tightest bounds in this low-energy regime. Recently, localized FRBs have also been utilized to constrain additional fundamental constants, such as the fine-structure constant and the proton-to-electron mass ratio \cite{2024MNRAS.533L..57K,2025JCAP...01..059L}.

Several recent studies have utilized FRBs with measured redshifts to constrain the Hubble constant ($H_0$), primarily leveraging Bayesian analysis techniques. Walters et al. \cite{2018ApJ...856...65W} pioneered the exploration of FRBs as cosmological probes by employing mock FRB catalogs, illustrating the potential for significant enhancements in the precision of cosmological parameter constraints when combined with data from the cosmic microwave background (CMB), baryonic acoustic oscillations (BAOs), supernovae, and $H_0$ measurements. Macquart et al. \cite{2020Natur.581..391M} used eight localized FRBs to constrain the baryonic matter density $\Omega_\mathrm{b} = 0.051^{+0.021}_{-0.025}h_{70}^{-1}$, where $h_{70} = H_0/(70 \rm\,km\,s^{-1}\,Mpc^{-1})$, consistent with prior estimates derived from CMB and Big Bang nucleosynthesis data. Additionally, a study employing strong lensing effects from 10 simulated FRBs estimated a value of $H_0 \approx 70 \rm\,km\,s^{-1}\,Mpc^{-1}$ \cite{2018NatCo...9.3833L}.

With the increasing number of localized FRBs, these constraints have undergone substantial refinements. Hagstotz et al. \cite{2022MNRAS.511..662H} analyzed nine localized FRBs, estimating $H_0 = 62.3\pm 9.1 \rm\,km\,s^{-1}\,Mpc^{-1}$, assuming a homogeneous contribution from the host galaxy, a factor that may limit the robustness of this estimate. Subsequently, Wu et al. \cite{2022MNRAS.515L...1W} classified 18 localized FRBs based on host galaxy morphology employing the IllustrisTNG simulation, to estimate individual host contributions, leading to an estimate of $H_0 = 68.81^{+4.99}_{-4.33}\rm\,km\,s^{-1}\,Mpc^{-1}$. However, the host DM values predicted by their model were found to be lower than the observed values for some FRBs. Furthermore, James et al. \cite{2022MNRAS.516.4862J} included 16 localized and 60 unlocalized FRBs detected by ASKAP, obtaining $H_0 = 73^{+12}_{-8}\rm\,km\,s^{-1}\,Mpc^{-1}$. The larger uncertainty in their result, despite the larger sample size, likely reflects the inclusion of systematic uncertainties. Subsequent research has investigated a variety of statistical methods, simulations, and FRB datasets to further constrain $H_0$ \cite{2024ApJ...965...57B,2023ApJ...946L..49L,2023ApJ...955..101W,2022arXiv221213433Z,2024MNRAS.527.7861G}. More recently, Fortunato et al. \cite{2025JCAP...01..018F} proposed a novel approach using artificial neural networks with 23 localized FRBs, yielding an estimate of $H_0 = 67.3 \pm 6.6 \rm\,km\,s^{-1}\,Mpc^{-1}$.

Although FRBs provide valuable preliminary constraints on cosmological parameters, we do not yet have enough observations with localizations to resolve the ongoing tensions in cosmology. While less stringent than those obtained from CMB and Type Ia supernovae (SNe\,Ia) data, the constraints derived from FRBs are tighter than those obtained from GW observations. The discrepancies in the reported values of $H_0$ primarily arise from uncertainties in modeling the contributions from the IGM and the host galaxy's DM. In this study, we apply various statistical methodologies to a comprehensive dataset of 64 localized, extragalactic FRBs to derive new constraints on $H_0$. Our analysis reveals substantial improvements in uncertainty compared to previous studies and shows consistency with recent late-time Universe measurements from the SH0ES collaboration \cite{2022ApJ...934L...7R}. These findings provide a significant contribution towards highlighting the tension between early- and late-time Universe $H_0$ estimates.

This article is structured as follows. In Section~\ref{Sec2}, we revisit the key relationships between the DM and host redshift, and their dependencies on $H_0$. Section~\ref{Sec3} describes our data sample, consisting of 64 localized FRBs, which is subsequently used to estimate $H_0$ through a Bayesian analysis. We employ three distinct methods with different likelihood functions to calculate $H_0$, accounting for the various uncertainties in the DM components of FRBs. Section~\ref{Sec4} provides a detailed discussion of the results showing our results are in alignment with the late-time Universe measurements of $H_0$. Finally, our conclusions are presented in Section~\ref{Sec5}.


\section{Revisiting dispersion measure of Fast Radio Bursts}\label{Sec2}

A key feature exhibited by FRBs is the dispersion sweep observable in the frequency-time domain. This effect arises from the presence of the ionized plasma along the line of sight of the radio waves from the source to the telescope. The degree of dispersion is characterized by DM, which quantifies the total column density of free electrons encountered by the signal. The DM is primarily composed of contributions from four distinct regions: the Milky Way Galaxy ($\mathrm{DM}_\mathrm{MW}$), its circumgalactic halo ($\mathrm{DM}_\mathrm{Halo}$), the IGM  ($\mathrm{DM}_\mathrm{IGM}$), and the host galaxy of the FRB ($\mathrm{DM}_\mathrm{Host}$). Denoting the source redshift as $z_\mathrm{S}$, the total DM contributions can be expressed as
\begin{align}\label{Eq: DM}
    \mathrm{DM} = \mathrm{DM}_\mathrm{MW} + \mathrm{DM}_\mathrm{Halo} + \mathrm{DM}_\mathrm{IGM}(z_\mathrm{S}) + \frac{\mathrm{DM}_\mathrm{Host}}{1+z_\mathrm{S}}.
\end{align}

Given our well-established understanding of the Galactic distribution of free electrons, $\mathrm{DM}_\mathrm{MW}$ is modeled with reasonable accuracy \cite{2002astro.ph..7156C,2017ApJ...835...29Y}. Additionally, it was already estimated that the Galactic halo contributes approximately $\mathrm{DM}_\mathrm{Halo}\approx 50-80\rm\,pc\,cm^{-3}$, independent of the contribution from the Galactic interstellar medium \cite{2019MNRAS.485..648P}. The remaining components, specifically $\mathrm{DM}_\mathrm{IGM}$ and $\mathrm{DM}_\mathrm{Host}$ remain less well constrained due to the challenges inherent in their measurements. However, simulations from IllustrisTNG suggest that for non-repeating FRBs, the median host galaxy contribution is $\mathrm{DM}_\mathrm{Host} = 33(1+z_\mathrm{S})^{0.84}\rm\,pc\,cm^{-3}$, whereas for repeating FRBs, it varies between $35(1+z_\mathrm{S})^{1.08}\rm\,pc\,cm^{-3}$ for dwarf galaxies and $96(1+z_\mathrm{S})^{0.83}\rm\,pc\,cm^{-3}$ for spiral galaxies \cite{2020ApJ...900..170Z}.

The average $\mathrm{DM}_\mathrm{IGM}$, known as the Macquart relation, is expressed as \cite{2020Natur.581..391M}
\begin{align}\label{Eq: Macquart}
    \langle \mathrm{DM}_\mathrm{IGM}(z_\mathrm{S})\rangle = \frac{3c \Omega_\mathrm{b} H_0^2}{8\pi G m_\mathrm{p}} \int_{0}^{z_\mathrm{S}} \frac{f_\mathrm{IGM}(z)\chi(z)(1 + z)}{H(z)} \dd{z},
\end{align}
where $c$ represents the speed of light, $G$ is the gravitational constant, $m_\mathrm{p}$ is the proton mass, $f_\mathrm{IGM}$ is the baryon fraction in the IGM, and $\chi(z)$ is the ionization fraction along the line of sight. The ionization fraction $\chi(z)$ is given by
\begin{align}
    \chi(z) = Y_\mathrm{H}\chi_{\mathrm{e},\mathrm{H}}(z) + \frac{1}{2}Y_\mathrm{p}\chi_{\mathrm{e},\mathrm{He}}(z), 
\end{align}
where $\chi_{\mathrm{e},\mathrm{H}}(z)$ and $\chi_{\mathrm{e},\mathrm{He}}(z)$ represent the ionization fractions of intergalactic hydrogen and helium, respectively, with mass fractions $Y_\mathrm{H} = 3/4$ and $Y_\mathrm{p} = 1/4$. For our calculations, we adopt $f_\mathrm{IGM} = 0.85$ following \cite{2022ApJ...931...88C}. The Hubble function $H(z)$ encodes the information of underlying cosmological model. Under the standard $\Lambda$ cold dark matter ($\Lambda$CDM) framework, neglecting contributions from radiation and curvature, it is given by $H(z) = H_0\sqrt{\Omega_\mathrm{m} \left(1+z\right)^3 + \Omega_\Lambda}$, where $\Omega_\mathrm{m}$ and $\Omega_\Lambda$ are the present-day matter and vacuum energy density fractions, respectively, constrained by the condition $\Omega_\mathrm{m} + \Omega_\Lambda = 1$.


\section{Data selection and Hubble constant estimation}\label{Sec3}

In this study, we analyze 64 FRBs that have been precisely localized within their host galaxies as of August 2024, which means their source redshifts ($z_\mathrm{S}$) have been measured. The relevant observables of these FRBs are presented in Table~\ref{Table: FRB}. Considering the NE2001 model \cite{2002astro.ph..7156C} of the Galactic distribution of free electrons, we present $\mathrm{DM}_\mathrm{MW}$ for each of the bursts. As discussed in the previous section, $\mathrm{DM}_\mathrm{Halo}$ is calculated based on the results of \cite{2019MNRAS.485..648P}. To account for the host galaxy’s contribution to DM, we categorize these FRBs into two groups. For FRBs with reported $\mathrm{DM}_\mathrm{Host}$ values, we directly use their maximum inferred values. For those without available $\mathrm{DM}_\mathrm{Host}$ data, we apply the model given in \cite{2020ApJ...900..170Z}, as described earlier, to estimate the missing values. By substituting all these values into Equation~\eqref{Eq: DM}, we calculate $\mathrm{DM}_\mathrm{IGM}$ for each FRB. 

\onecolumn
\begin{longtable}{|l|l|l|l|l|l|l|}
\caption{List of 64 localized FRBs as of August 2024 in chronological order. FRBs in bold indicate that their $\mathrm{DM}_\mathrm{Host}$ values are reported. $\mathrm{DM}_\mathrm{MW}$ is calculated based on NE2001 model.}
\label{Table: FRB}\\
\hline
Name & $\mathrm{DM}_\mathrm{obs}$ & $\mathrm{DM}_\mathrm{MW}$ & $z_\mathrm{S}$ & Repeater & Ref. \\
& $(\rm pc\,cm^{-3})$ & $(\rm pc\,cm^{-3})$  & & (Y/N) & \\
\hline
\textbf{FRB\,20121102A} & $557.0\pm2.0$ & 188.0 & 0.19273 & Y & \cite{2017ApJ...834L...7T}\\
FRB\,20171020A & $114.1\pm0.2$ & 38.0 & 0.0086 & N & \cite{2018ApJ...867L..10M} \\
\textbf{FRB\,20180301A} & $522.0\pm0.2$ & 152.0 & 0.3304 & Y & \cite{2019MNRAS.486.3636P}\\
\textbf{FRB\,20180916B} & $349.3\pm0.2$ & 200.0 & 0.0337 & Y & \cite{2020Natur.577..190M}\\
\textbf{FRB\,20180924B} & $361.42\pm0.06$ & 40.5 & 0.3214 & N & \cite{2019Sci...365..565B}\\
\textbf{FRB\,20181112A} & $589.27\pm0.03$ & 102.0 & 0.4755 & N & \cite{2019Sci...366..231P}\\
FRB\,20181220A & $209.4\pm0.1$ & 126.0 & 0.02746 & N & \cite{2024ApJ...971L..51B}\\
FRB\,20181223C & $112.5\pm0.1$ & 20.0 & 0.03024 & N & \cite{2024ApJ...971L..51B}\\
\textbf{FRB\,20190102C} & $363.6\pm0.3$ & 57.3 & 0.291 & N & \cite{2020Natur.581..391M}\\
FRB\,20190110C & $221.6\pm1.6$ & 37.1 & 0.12244 & Y & \cite{2024ApJ...961...99I}\\
FRB\,20190303A & $222.4\pm0.7$ & 29.0 & 0.064 & Y & \cite{2023ApJ...950..134M}\\
FRB\,20190418A & $184.5\pm0.1$ & 71.0 & 0.07132 & N & \cite{2024ApJ...971L..51B}\\
FRB\,20190425A & $128.2\pm0.1$ & 49.0 & 0.03122 & N & \cite{2024ApJ...971L..51B}\\
\textbf{FRB\,20190520B} & $1204.7\pm10.0$ & 113 & 0.241 & Y & \cite{2022ApJ...931...87O}\\
\textbf{FRB\,20190523A} & $760.8\pm0.6$ & 37.0 & 0.66 & N & \cite{2019ApJ...886..135P}\\
\textbf{FRB\,20190608B} & $338.7\pm0.5$ & 37.2 & 0.1178 & N & \cite{2020ApJ...901..134S}\\
FRB\,20190611B & $321.4\pm0.2$ & 57.83 & 0.3778 & N & \cite{2020ApJ...903..152H}\\
\textbf{FRB\,20190614D} & $959.2\pm5.0$ & 83.5 & 0.6 & N & \cite{2020ApJ...899..161L}\\
FRB\,20190711A & $593.1\pm0.4$ & 56.4 & 0.522 & Y & \cite{2021MNRAS.500.2525K}\\
FRB\,20190714A & $504.0\pm2.0$ & 39.0 & 0.2365 & N & \cite{2020ApJ...903..152H}\\
\textbf{FRB\,20191001A} & $506.92\pm0.04$ & 44.7 & 0.234 & N & \cite{2022MNRAS.512L...1K}\\
FRB\,20191106C & $332.2\pm0.7$ & 25.0 & 0.10775 & Y & \cite{2024ApJ...961...99I}\\
FRB\,20191228A & $297.50\pm0.05$ & 33.0 & 0.2432 & N & \cite{2022AJ....163...69B}\\
FRB\,20200223B & $201.8\pm0.4$ & 45.6 & 0.0602 & Y & \cite{2024ApJ...961...99I}\\
FRB\,20200430A & $380.1\pm0.4$ & 27.0 & 0.16 & N & \cite{2020ApJ...903..152H}\\
FRB\,20200906A & $577.80\pm0.02$ & 36.0 & 0.3688 & N & \cite{2022AJ....163...69B}\\
FRB\,20201123A & $433.55\pm0.36$ & 251.93 & 0.05 & N & \cite{2022MNRAS.514.1961R}\\
\textbf{FRB\,20201124A} & $413.52\pm0.05$ & 123.2 & 0.098 & Y & \cite{2022MNRAS.513..982R}\\
FRB\,20210117A & $730.0\pm1.0$ & 34.4 & 0.2145 & N & \cite{2023ApJ...948...67B}\\
FRB\,20210320C & $384.8\pm0.3$ & 42.0 & 0.2797 & N & \cite{2022MNRAS.516.4862J}\\
FRB\,20210405I & $565.17\pm0.49$ & 516.1 & 0.066 & N & \cite{2024MNRAS.527.3659D}\\
\textbf{FRB\,20210410D} & $578.78\pm2.00$ & 56.2 & 0.1415 & N & \cite{2023MNRAS.524.2064C}\\
\textbf{FRB\,20210603A} & $500.147\pm0.004$ & 40.0 & 0.177 & N & \cite{2024NatAs...8.1429C}\\
FRB\,20210807D & $251.9\pm0.2$ & 121.2 & 0.12927 & N & \cite{2022MNRAS.516.4862J}\\
FRB\,20211127I & $234.83\pm0.08$ & 42.5 & 0.0469 & N & \cite{2022MNRAS.516.4862J}\\
\textbf{FRB\,20211203C} & $636.2\pm0.4$ & 63.0 & 0.3437 & N & \cite{2024arXiv240802083S}\\
FRB\,20211212A & $206.0\pm5.0$ & 27.1 & 0.0715 & N & \cite{2022MNRAS.516.4862J}\\
FRB\,20220105A & $583.0\pm2.0$ & 22.0 & 0.2785 & N & \cite{2024arXiv240802083S}\\
FRB\,20220207C & $262.38\pm0.01$ & 79.3 & 0.04304 & N & \cite{2024ApJ...967...29L}\\
FRB\,20220307B & $499.27\pm0.06$ & 135.7 & 0.248123 & N & \cite{2024ApJ...967...29L}\\
FRB\,20220310F & $462.240\pm0.005$ & 45.4 & 0.477958 & N & \cite{2024ApJ...967...29L}\\
FRB\,20220319D & $110.95\pm0.02$ & 65.25 & 0.011 & N & \cite{2023arXiv230101000R}\\
FRB\,20220418A & $623.25\pm0.01$ & 37.6 & 0.622 & N & \cite{2024ApJ...967...29L}\\
FRB\,20220501C & $449.5\pm0.2$ & 31.0 & 0.381 & N & \cite{2024arXiv240802083S}\\
FRB\,20220506D & $396.97\pm0.02$ & 89.1 & 0.30039 & N & \cite{2024ApJ...967...29L}\\
FRB\,20220509G & $269.53\pm0.02$ & 55.2 & 0.0894 & N & \cite{2024ApJ...967...29L}\\
\textbf{FRB\,20220610A} & $1458.0\pm0.2$ & 31.0 & 1.017 & N & \cite{2024ApJ...963L..34G}\\
FRB\,20220725A & $290.4\pm0.3$ & 31.0 & 0.1926 & N & \cite{2024arXiv240802083S}\\
FRB\,20220825A & $651.24\pm0.06$ & 79.7 & 0.241397 & N & \cite{2024ApJ...967...29L}\\
FRB\,20220912A & $219.46\pm0.04$ & 125.0 & 0.077 & Y & \cite{2023ApJ...949L...3R}\\
FRB\,20220914A & $631.28\pm0.04$ & 55.2 & 0.1139 & N & \cite{2024ApJ...967...29L}\\
FRB\,20220918A & $656.8\pm0.4$ & 41.0 & 0.491 & N & \cite{2024arXiv240802083S}\\
FRB\,20220920A & $314.99\pm0.01$ & 40.3 & 0.158239 & N & \cite{2024ApJ...967...29L}\\
FRB\,20221012A & $441.08\pm0.70$ & 54.4 & 0.284669 & N & \cite{2024ApJ...967...29L}\\
FRB\,20221106A & $343.8\pm0.8$ & 35.0 & 0.2044 & N & \cite{2024arXiv240802083S}\\
FRB\,20230526A & $361.4\pm0.2$ & 50.0 & 0.157 & N & \cite{2024arXiv240802083S}\\
FRB\,20230708A & $411.51\pm0.05$ & 50.0 & 0.105 & N & \cite{2024arXiv240802083S}\\
FRB\,20230718A & $476.6\pm0.5$ & 393.0 & 0.0357 & N & \cite{2024ApJ...962L..13G}\\
FRB\,20230902A & $440.1\pm0.1$ & 34.0 & 0.3619 & N & \cite{2024arXiv240802083S}\\
FRB\,20231226A & $329.9\pm0.1$ & 145.0 & 0.1569 & N & \cite{2024arXiv240802083S}\\
\textbf{FRB\,20240114A} & $527.7\pm0.1$ & 40.0 & 0.42 & Y & \cite{2024ApJ...977..177K}\\
FRB\,20240201A & $374.5\pm0.2$ & 38.0 & 0.042729 & N & \cite{2024arXiv240802083S}\\
FRB\,20240210A & $283.73\pm0.05$ & 31.0 & 0.023686 & N & \cite{2024arXiv240802083S}\\
FRB\,20240310A & $601.8\pm0.2$ & 36.0 & 0.127 & N & \cite{2024arXiv240802083S}\\
\hline
\end{longtable}

\twocolumn

\begin{figure}[htpb]
    \centering
    \includegraphics[scale=0.5]{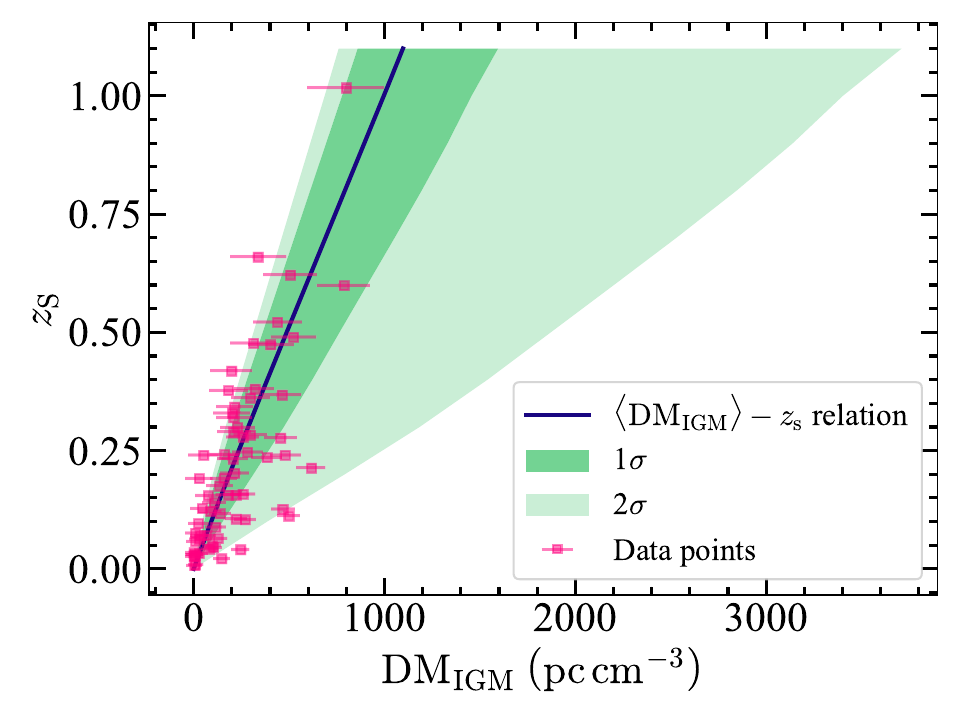}
    \caption{The host redshift values for the localized FRBs are plotted against their estimated DM contribution from IGM along with their error bars. The black solid line depicts the $\langle \mathrm{DM}_\mathrm{IGM}(z_\mathrm{S})\rangle$ as a function of $z_\mathrm{S}$ assuming $\Lambda$CDM cosmology with $H_0=73\rm\,km\,s^{-1}\,Mpc^{-1}$, $\Omega_\mathrm{m} = 0.30966$, $\Omega_\Lambda = 1-\Omega_\mathrm{m}$, and $\Omega_\mathrm{b}=0.04897$. Green shaded regions represent 1$\sigma$ and 2$\sigma$ confidence regions of $\mathrm{DM}_\mathrm{IGM}$.}
    \label{Fig: DM_z}
\end{figure}

Figure~\ref{Fig: DM_z} illustrates the $z_\mathrm{S}$ values of these FRBs plotted against their corresponding $\mathrm{DM}_\mathrm{IGM}$ values, including error bars. The figure also shows the expected average $\langle \mathrm{DM}_\mathrm{IGM}(z_\mathrm{S})\rangle$ under the standard $\Lambda$CDM cosmology with $H_0 = 73\rm\,km\,s^{-1}\,Mpc^{-1}$, $\Omega_\mathrm{m} = 0.30966$, $\Omega_\Lambda = 1-\Omega_\mathrm{m}$, and $\Omega_\mathrm{b}=0.04897$. Additionally, the 1$\sigma$ and 2$\sigma$ confidence regions for $\langle \mathrm{DM}_\mathrm{IGM}(z_\mathrm{S})\rangle$, derived from the results of \cite{2019MNRAS.484.1637J} using the Illustris simulation, are depicted. These regions account for the inhomogeneous distribution of ionized gas in the IGM.

We now present our results on constraining $H_0$ using the analyzed data sample. The likelihood analysis is conducted within the Bayesian framework, utilizing the Markov chain Monte Carlo (MCMC) method in Python \cite{2013PASP..125..306F}. In particular, we select three distinct likelihood functions for this study to explore different modeling approaches.


\subsection{Likelihood 1: incorporating uncertainties in $\mathrm{DM}_\mathrm{IGM}$ measurements}

We assume that the individual likelihood of $\mathrm{DM}_\mathrm{IGM}$ for each FRB follows a Gaussian distribution, given by
\begin{align}\label{Eq: chisq}
    P_i\left(\mathrm{DM}_{\mathrm{IGM},i}\mid z_{\mathrm{S},i}\right) = \frac{1}{\sqrt{2\pi\sigma_i^2}} e^{-\frac{\left(\langle\mathrm{DM}_\mathrm{IGM}\left(z_{\mathrm{S},i}\right)\rangle-\mathrm{DM}_{\mathrm{IGM},i}\right)^2}{2\sigma_i^2}}.
\end{align}
where
\begin{align}
    \sigma_i^2 = \sigma_{\mathrm{obs},i}^2 + \sigma_{\mathrm{MW},i}^2 + \sigma_{\mathrm{Halo},i}^2 + \sigma_{\mathrm{IGM},i}^2 + \frac{\sigma_{\mathrm{Host},i}^2}{1+z_{\mathrm{S},i}}.
\end{align}
Here $\sigma_\mathrm{obs}$ is the error bar in the observed DM  of each FRB, included in Table~\ref{Table: FRB} and rests are error bars corresponding to each DM components. Thereby we define the joint likelihood function as
\begin{align}\label{Eq: Likelihood1}
    \mathcal{L} = \prod_{i=1}^{N_\mathrm{FRB}} P_i\left(\mathrm{DM}_{\mathrm{IGM},i}\mid z_{\mathrm{S},i}\right),
\end{align}
where $N_\mathrm{FRB}$ is the total number of FRBs in the dataset. To obtain constraints on $H_0$, the joint likelihood is maximized with respect to $H_0$. As discussed previously, we adopt $\mathrm{DM}_\mathrm{MW}$ values from NE2001 model, while its error bar is considered as $\sigma_\mathrm{MW}=30\rm\,pc\,cm^{-3}$ \cite{2005AJ....129.1993M}. For the Galactic halo, as it contributes $\mathrm{DM}_\mathrm{Halo}\approx 50-80\rm\,pc\,cm^{-3}$, we draw values for each FRB from a uniform distribution $\mathcal{U}[50-80] \rm\,pc\,cm^{-3}$, and hence it indicates $\sigma_\mathrm{Halo}=\sqrt{75}\rm\,pc\,cm^{-3}$. For host galaxy contributions, $\mathrm{DM}_\mathrm{Host}$ is considered either from reported values or, if unavailable, from IllustrisTNG simulations, while we assume large scatter by considering $\sigma_\mathrm{Host}=50\rm\,pc\,cm^{-3}$ for all FRBs. The error bars in IGM component is considered from the simulated results in \cite{2019MNRAS.484.1637J} utilizing inhomogeneous electron distribution.

\begin{figure}[htpb]
    \centering
    \includegraphics[scale=0.5]{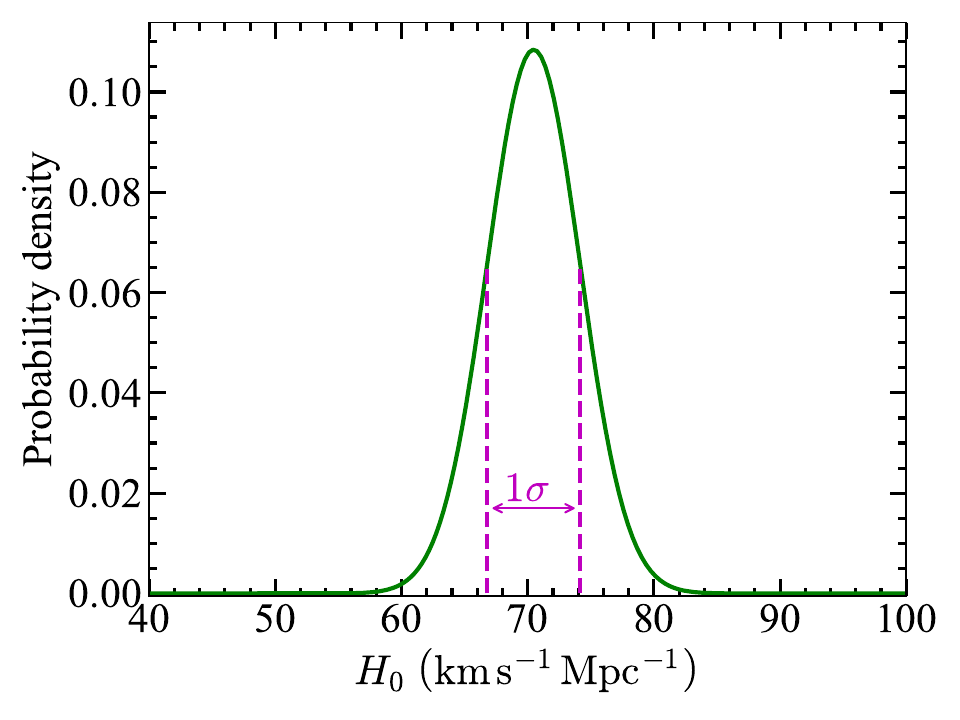}
    \caption{Probability distribution of joint likelihood function of Equation~\eqref{Eq: Likelihood1} with respect to $H_0$ along with its 1$\sigma$ confidence interval. The likelihood is maximized for $H_0 = 70.42^{+3.73}_{-3.63}\rm\,km\,s^{-1}\,Mpc^{-1}$.}
    \label{Fig: Hubble1}
\end{figure}

Figure~\ref{Fig: Hubble1} presents the probability density function of the joint likelihood for all 64 localized FRBs as a function of $H_0$, along with the $1\sigma$ confidence region within the framework of $\Lambda$CDM cosmology. The likelihood is maximized at $H_0 = 70.42^{+3.73}_{-3.63}\rm\,km\,s^{-1}\,Mpc^{-1}$, within the $1\sigma$ confidence level. This value represents the most robust constraint on $H_0$ derived from localized FRB data by comparing the calculated $\mathrm{DM}_\mathrm{IGM}$ values with the mean $\langle\mathrm{DM}_\mathrm{IGM}\rangle$, based on the Macquart relation.

\begin{figure}[htpb]
    \centering
    \includegraphics[scale=1.1]{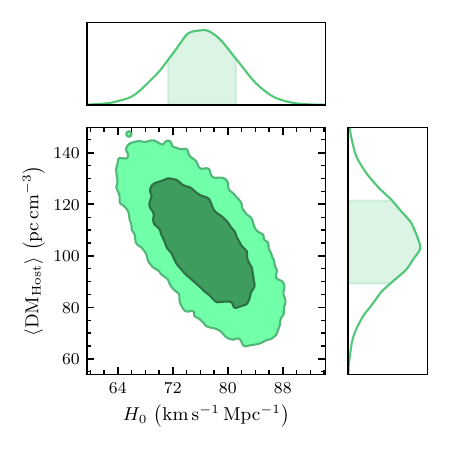}
    \caption{1$\sigma$ and 2$\sigma$ confidence regions of the joint likelihood of Equation~\eqref{Eq: Likelihood1} when it optimized over both $H_0$ and $\langle\mathrm{DM}_\mathrm{Host}\rangle$. This likelihood is maximized for $H_0 = 76.26^{+4.86}_{-4.85}\rm\,km\,s^{-1}\,Mpc^{-1}$ and $\langle\mathrm{DM}_\mathrm{Host}\rangle = 105.15^{+16.15}_{-15.77}\rm\,pc\,cm^{-3}$. This plot is prepared with ChainConsumer function of Python \cite{Hinton2016}.}
    \label{Fig: Hubble+Host}
\end{figure}

This estimate of $H_0$ is limited by the uncertainty in the host galaxy DM contributions. As previously mentioned, $\mathrm{DM}_\mathrm{Host}$ of each FRB is either taken from its maximum reported value or from IllustrisTNG simulation, with the latter typically yielding lower values. It is currently impossible to accurately measure $\mathrm{DM}_\mathrm{Host}$ for individual FRBs; however, its mean value, $\langle\mathrm{DM}_\mathrm{Host}\rangle$, can be estimated through MCMC analysis. By treating both $H_0$ and $\langle\mathrm{DM}_\mathrm{Host}\rangle$ as free parameters, we maximize the joint likelihood from Equation~\eqref{Eq: Likelihood1}. Figure~\ref{Fig: Hubble+Host} shows the joint likelihood distribution, revealing a mean host DM contribution of approximately $105.15^{+16.15}_{-15.77}\rm\,pc\,cm^{-3}$ and a revised Hubble constant of $H_0 = 76.26^{+4.86}_{-4.85}\rm\,km\,s^{-1}\,Mpc^{-1}$. The error bars increase in this scenario, but it is noteworthy that this estimated $H_0$ aligns with values derived from late-time Universe supernovae data.


\subsection{Likelihood 2: incorporating uncertainties in $\mathrm{DM}_\mathrm{IGM}$ and $\mathrm{DM}_\mathrm{Host}$ measurements}

Given the substantial uncertainty in the host DM contribution, previously reported $H_0$ values should be interpreted with caution. To mitigate this uncertainty, we introduce a new likelihood function incorporating uncertainties in both $\mathrm{DM}_\mathrm{IGM}$ and $\mathrm{DM}_\mathrm{Host}$. If $X(x)$, $Y(y)$, and $Z(z)$ are random variables ($x,y,z\geq0$) with probability distributions $P_X(x)$, $P_Y(y)$, and $P_Z(z)$, respectively, such that $z=x+y$, then $P_Z(z) = \int_0^z P_X(x) P_Y(z-x) \dd{x}$. Applying this mathematical identity, we define the excess DM as $\mathrm{DM}_\mathrm{exc} = \mathrm{DM} - \mathrm{DM}_\mathrm{MW} - \mathrm{DM}_\mathrm{Halo} = \mathrm{DM}_\mathrm{IGM} + \mathrm{DM}_\mathrm{Host}/(1+z_\mathrm{S})$. We adopt the NE2001 model for $\mathrm{DM}_\mathrm{MW}$ and a uniform distribution $\mathcal{U}[50-80] \rm\,pc\,cm^{-3}$ for $\mathrm{DM}_\mathrm{Halo}$ to construct the following joint likelihood function \cite{2020Natur.581..391M}
\begin{align}\label{Eq: Likelihood2}
    \mathcal{L} = \prod_{i=1}^{N_\mathrm{FRB}} P_i\left(\mathrm{DM}_{\mathrm{exc},i} \mid z_{\mathrm{S},i}\right),
\end{align}
which requires maximization over $H_0$. Here
\begin{align}
     P_i\left(\mathrm{DM}_{\mathrm{exc},i} \mid z_{\mathrm{S},i}\right) &= \int_0^{\mathrm{DM}_{\mathrm{exc},i}} P_\mathrm{Host}\left(\frac{\mathrm{DM}_\mathrm{Host}}{1+z_{\mathrm{S},i}}\right) \nonumber \\ &\times P_\mathrm{IGM}\left(\mathrm{DM}_{\mathrm{exc},i}-\frac{\mathrm{DM}_\mathrm{Host}}{1+z_{\mathrm{S},i}} \right)\dd{\mathrm{DM}_\mathrm{Host}}.
\end{align}
Previous studies suggest that $\mathrm{DM}_\mathrm{IGM}$ follows a specific probability distribution, given by \cite{2020Natur.581..391M}
\begin{align}
    P_\mathrm{IGM}\left(\Delta_\mathrm{IGM}\right) = A \Delta_\mathrm{IGM}^{-\beta_2} e^{-\frac{\left(\Delta_\mathrm{IGM}^{-\beta_1}-C_0\right)^2}{2\beta_1^2\sigma_\mathrm{DM}^2}},
\end{align}
where $\Delta_\mathrm{IGM} = \mathrm{DM}_\mathrm{IGM}/\langle\mathrm{DM}_\mathrm{IGM}\rangle$, $\sigma_\mathrm{DM}$ is its standard deviation, $A$, $\beta_1$, $\beta_2$, and $C_0$ are model parameters. As noted in \cite{2020Natur.581..391M}, the form provides a good fit to the observational data within their semi-analytic models and hydrodynamic simulations across a range of halo masses and redshifts. This distribution asymptotically approaches a Gaussian in the limit of small $\sigma_\text{DM}$. We adopt the parameter values from \cite{2021ApJ...906...49Z}, based on the IllustrisTNG simulation. The inherent difficulty in measuring $\mathrm{DM}_\mathrm{Host}$ introduces significant uncertainty regarding its exact distribution. Following \cite{2020Natur.581..391M}, we employ a log-normal probability distribution, given by
\begin{align}
    P_\mathrm{Host}\left(\mathrm{DM}_\mathrm{Host}\right) = \frac{1}{\sqrt{2\pi}\mathrm{DM}_\mathrm{Host}\sigma_\mathrm{Host}} e^{-\frac{\left(\ln{\mathrm{DM}_\mathrm{Host}}-\mu_\mathrm{Host}\right)^2}{2\sigma_\mathrm{Host}^2}},
\end{align}
with $e^{\mu_\mathrm{Host}}$ being the median and $\left(e^{\sigma_\mathrm{Host}^2}-1\right)e^{2\mu_\mathrm{Host}+\sigma_\mathrm{Host}^2}$ its variance. This choice enables to capture high $\text{DM}_\text{Host}$ values that may result from local environments of the FRB, such as HII regions or circumstellar media. Our analysis considers four different combinations of $\mu_\mathrm{Host}$ and $\sigma_\mathrm{Host}$, as listed in Table~\ref{Table: mu+sigma}. These distribution functions effectively capture the uncertainties associated with $\mathrm{DM}_\mathrm{IGM}$ and $\mathrm{DM}_\mathrm{Host}$ values. Due to the precise localization of these FRBs, their redshift values are subject to minimal uncertainty.

\begin{table}[htpb]
    \centering
    \caption{Parameters involved in host DM probability distribution along with the estimated Hubble constant value in each case.}
    \label{Table: mu+sigma}
    \resizebox{\columnwidth}{!}{
    \begin{tabular}{|l|l|l|l|l|l|}
    \hline
        Model & $e^{\mu_\mathrm{Host}}$ & $\sigma_\mathrm{Host}$ & Ref. & $H_0$ (Likelihood 2) & $H_0$ (Likelihood 3)\\
        & & & & ($\rm km\,s^{-1}\,Mpc^{-1}$) & ($\rm km\,s^{-1}\,Mpc^{-1}$) \\
        \hline
        Model I & 66 & 0.42 & \cite{2020Natur.581..391M} & $74.77_{-2.65}^{+2.51}$ & $77.23_{-2.71}^{+2.62}$ \\
        Model II & 30 & 0.17 & \cite{2020MNRAS.494..665L} & $73.41_{-2.71}^{+2.28}$ & $76.52_{-2.57}^{+2.65}$ \\
        Model III & 130 & 0.53 & \cite{2022MNRAS.516.4862J} & $73.74_{-2.77}^{+2.43}$ & $76.16_{-2.89}^{+2.48}$ \\
        Model IV & $f(z_\mathrm{S})$* & $f(z_\mathrm{S})$* & \cite{2020ApJ...900..170Z} & $83.53_{-2.92}^{+2.32}$ & $85.08_{-2.66}^{+2.71}$\\
        \hline
    \end{tabular}
    }
    \begin{tablenotes}
      \item *In model IV, $e^{\mu_\mathrm{Host}}$ and $\sigma_\mathrm{Host}$ change with the host redshift; hence we do not mention any values in the table.
    \end{tablenotes}
\end{table}

\begin{figure}[htpb]
    \centering
    \includegraphics[scale=0.5]{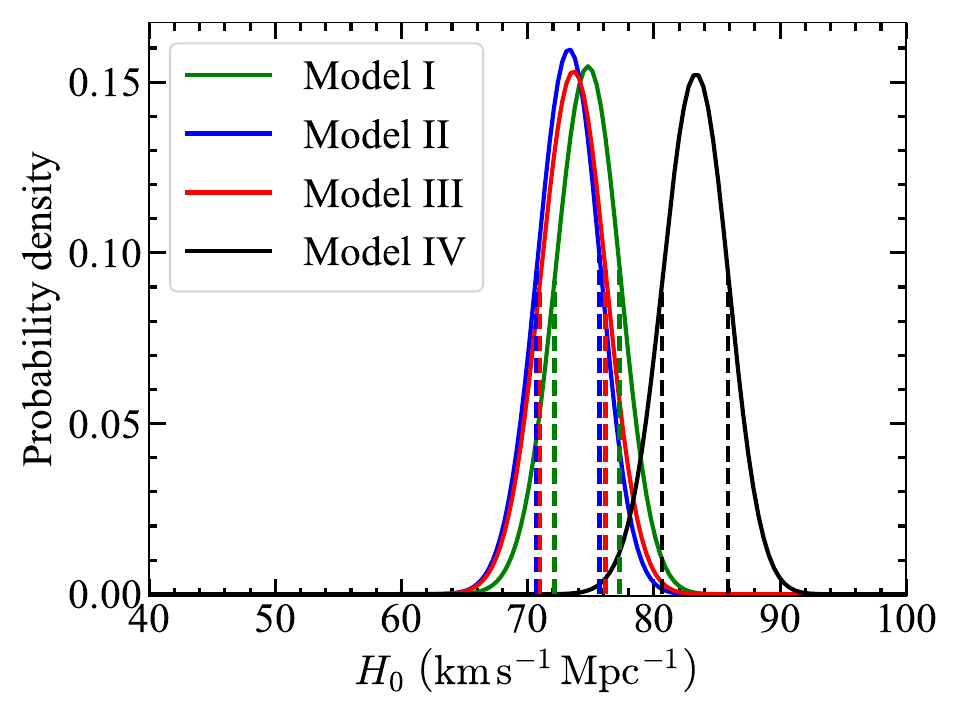}
    \caption{Probability distribution of joint likelihood function of Equation~\eqref{Eq: Likelihood2} with respect to $H_0$ along with its 1$\sigma$ confidence interval for four different host DM models.}
    \label{Fig: Hubble2}
\end{figure}

Table~\ref{Table: mu+sigma} presents the values of $H_0$ with 1$\sigma$ uncertainties for the four Host DM models. Figure~\ref{Fig: Hubble2} illustrates the probability distribution of the joint likelihood function of Equation~\eqref{Eq: Likelihood2} with respect to $H_0$, along with its 1$\sigma$ confidence interval for each host DM model. Except for Model IV, the other three models yield similar $H_0$. Notably, the error bars decrease to approximately $\pm2.5\rm\,km\,s^{-1}\,Mpc^{-1}$ compared to those obtained using likelihood 1 in the aforementioned section. All these $H_0$ values are consistent with the SH0ES collaboration result for nearby SNe\,Ia \cite{2022ApJ...934L...7R} and the reduced 1$\sigma$ uncertainty error bars no longer overlap with $H_0$ value derived from CMB data.

\begin{figure}[htpb]
    \centering
    \includegraphics[scale=0.5]{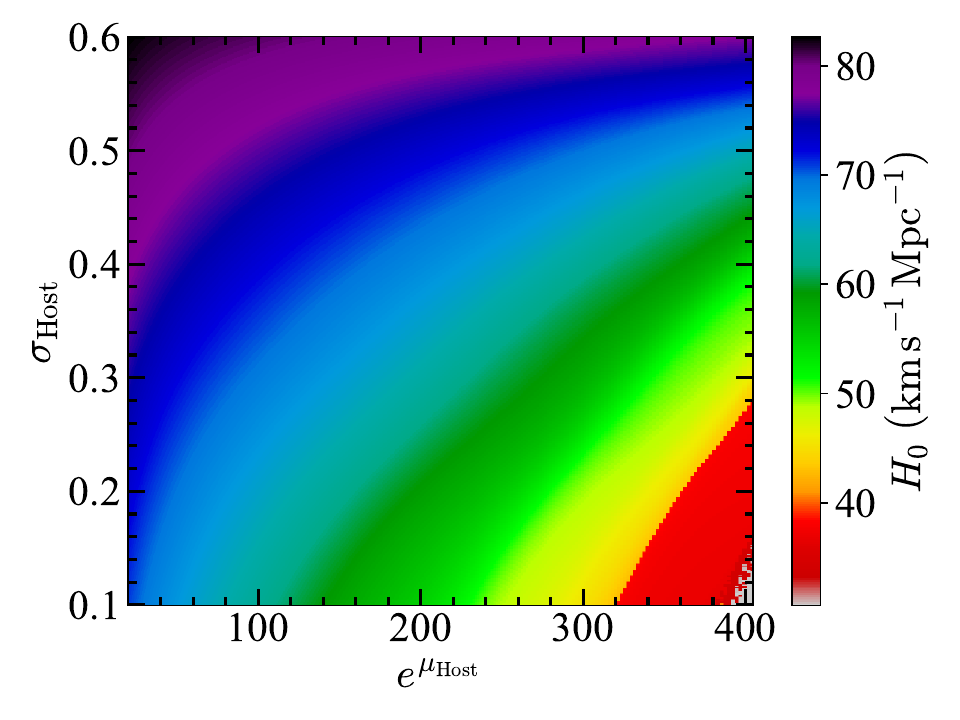}
    \caption{Values of Hubble constant, indicated by the color bar, estimated using likelihood of Equation~\eqref{Eq: Likelihood2} for different combinations of $e^{\mu_\mathrm{Host}}$ and $\sigma_\mathrm{Host}$ in host DM model.}
    \label{Fig: H0+mu+sigma}
\end{figure}

It is important to note that FRB data alone cannot definitively predict the value of $H_0$. This limitation primarily stems from the significant uncertainties in $\mathrm{DM}_\mathrm{IGM}$ and $\mathrm{DM}_\mathrm{Host}$. As depicted in Figure~\ref{Fig: Hubble2}, different host DM models, characterized by varying combinations of $e^{\mu_\mathrm{Host}}$ and $\sigma_\mathrm{Host}$, produce distinct results. Therefore, host DM modeling plays a crucial role in determining the $H_0$ value. By exploring numerous combinations of these two model parameters, we plot the corresponding $H_0$ values in Figure~\ref{Fig: H0+mu+sigma}. The results demonstrate that various combinations of $e^{\mu_\mathrm{Host}}$ and $\sigma_\mathrm{Host}$ can yield same $H_0$. Furthermore, we observe that a scenario with a high host DM component and low scatter in their DM values, represented by high $e^{\mu_\mathrm{Host}}$ and low $\sigma_\mathrm{Host}$, leads to very low $H_0$ estimates. Such a scenario is highly unrealistic, as FRBs originate from diverse environments, including galaxies with varying morphologies, star formation rates, and other properties. Thus, in general, we can expect to have large $\sigma_\mathrm{Host}$ in their distribution. Conversely, if $\sigma_\mathrm{Host}$ is reasonably high, $H_0$ is estimated to be around or above $70\rm\,km\,s^{-1}\,Mpc^{-1}$, which is consistent with values obtained from different realistic simulations as listed in Table~\ref{Table: mu+sigma}. These results also depend on the choice of distribution function, especially both the functional form and the specific parameter values. For instance, adopting an alternative right-skewed distribution for $\text{DM}_\text{Host}$, with parameters chosen to yield similar probability densities up to $z_\mathrm{S} \sim 1$, does not significantly affect the overall results. However, as illustrated in Figure~\ref{Fig: H0+mu+sigma}, substantial variation in the distribution parameters, even for the same functional form, can lead to noticeable differences in the outcomes.


\subsection{Likelihood 3: incorporating uncertainties in $\mathrm{DM}_\mathrm{IGM}$, $\mathrm{DM}_\mathrm{Host}$, and $\mathrm{DM}_\mathrm{Halo}$ measurements}

While the Milky Way's DM can be reasonably modeled using either NE2001 \cite{2002astro.ph..7156C} or YMW16 \cite{2017ApJ...835...29Y} models, the values of the halo DM contribution remain a bit uncertain. Although previous studies have suggested $\mathrm{DM}_\mathrm{Halo}\approx 50-80\rm\,pc\,cm^{-3}$ \cite{2019MNRAS.485..648P}, the exact value for each FRB is unknown. Unlike the uncertainty in host DM contributions, the scatter in $\mathrm{DM}_\mathrm{Halo}$ is relatively small, and we anticipate minor improvements in results due to this variation. We now investigate the impact of the $\mathrm{DM}_\mathrm{Halo}$ uncertainty on the $H_0$ estimation. 

\begin{figure}[htpb]
    \centering
    \includegraphics[scale=0.5]{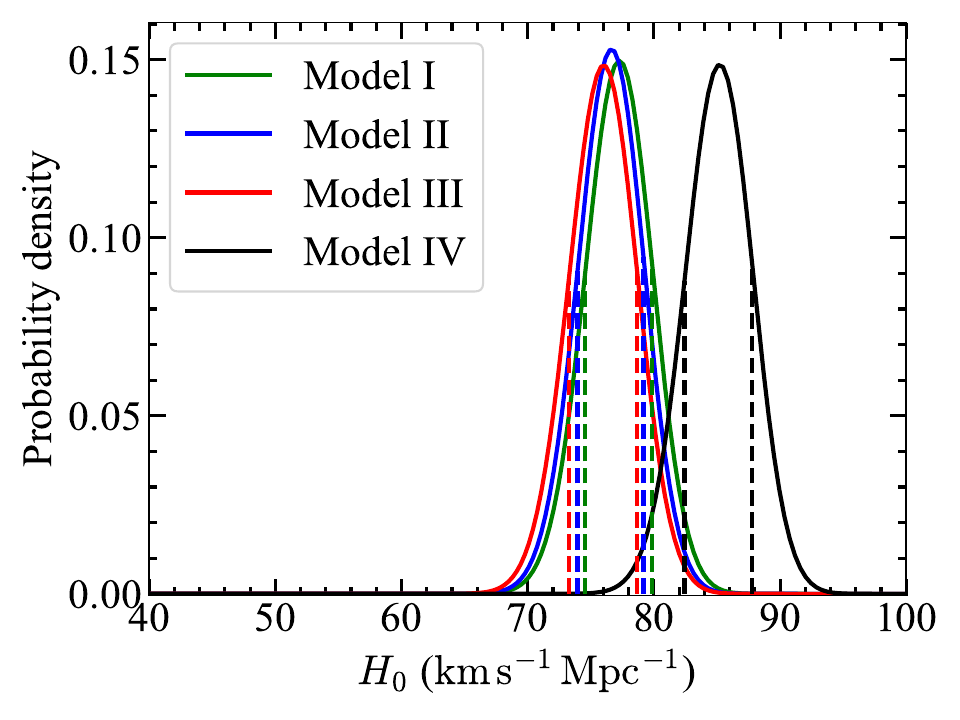}
    \caption{Probability distribution of joint likelihood function of Equation~\eqref{Eq: Likelihood3} with respect to $H_0$ together with its 1$\sigma$ confidence interval for four different host DM models.}
    \label{Fig: Hubble3}
\end{figure}

Defining $\mathrm{DM}' = \mathrm{DM} - \mathrm{DM}_\mathrm{MW} = \mathrm{DM}_\mathrm{exc} + \mathrm{DM}_\mathrm{Halo}$, we take into account this uncertainty in the modified likelihood function, given by
\begin{align}\label{Eq: Likelihood3}
    \mathcal{L} = \prod_{i=1}^{N_\mathrm{FRB}} P_i\left(\mathrm{DM}'_{i} \mid z_{\mathrm{S},i}\right),
\end{align}
where
\begin{align}
     P_i\left(\mathrm{DM}'_{i} \mid z_{\mathrm{S},i}\right) &= \int_0^{\mathrm{DM}'_{i}} \int_0^{\mathrm{DM}'_{i} - \mathrm{DM}_\mathrm{Halo}}  P_\mathrm{Halo}\left(\mathrm{DM}_{\mathrm{Halo}}\right)\nonumber \\ &\times  P_\mathrm{IGM}\left(\mathrm{DM}'_{i} - \frac{\mathrm{DM}_\mathrm{Host}}{1+z_{\mathrm{S},i}} - \mathrm{DM}_\mathrm{Halo} \right) \nonumber\\ &\times P_\mathrm{Host}\left(\frac{\mathrm{DM}_\mathrm{Host}}{1+z_{\mathrm{S},i}}\right) \dd{\mathrm{DM}_\mathrm{Halo}} \dd{\mathrm{DM}_\mathrm{Host}}.
\end{align}

We model the probability distribution of $\mathrm{DM}_\mathrm{Halo}$, $P_\mathrm{Halo}\left(\mathrm{DM}_{\mathrm{Halo}}\right)$, as a Gaussian distribution with mean $\mu_\mathrm{Halo}$ and standard deviation $\sigma_\mathrm{Halo}$, as follows:
\begin{align}
     P_\mathrm{Halo}\left(\mathrm{DM}_{\mathrm{Halo}}\right) = \frac{1}{\sqrt{2\pi\sigma_\mathrm{Halo}^2}}e^{-\frac{\left({\mathrm{DM}_\mathrm{Halo}}-\mu_\mathrm{Halo}\right)^2}{2\sigma_\mathrm{Halo}^2}}.
\end{align}
Assuming $\mu_\mathrm{Halo} \approx 65\rm\,pc\,cm^{-3}$ and $\sigma_\mathrm{Halo}\approx5\rm\,pc\,cm^{-3}$, we can mimic the cumulative function of this Gaussian distribution with the previously assumed uniform distribution $\mathcal{U}[50-80] \rm\,pc\,cm^{-3}$. Figure~\ref{Fig: Hubble3} illustrates the probability densities as a function of $H_0$, along with 1$\sigma$ confidence intervals for all four host DM models discussed in the previous section. The $H_0$ values at which the distributions are maximized are listed in the last column of Table~\ref{Table: mu+sigma}. As expected, the values exhibit slight deviations with the introduction of $\mathrm{DM}_\mathrm{Halo}$ uncertainty, although there is no significant improvement in error bars.

\section{Discussion}\label{Sec4}

One of the most pressing challenges in modern cosmology is the Hubble tension, which refers to the discrepancy between two independent sets of measurements of the Universe's current expansion rate, yielding different values for $H_0$. The first method relies on observations of the CMB as measured by missions like Planck. These early-Universe observations, combined with the $\Lambda$CDM model, predict a value of $H_0=67.36\pm0.54\rm\,km\,s^{-1}\,Mpc^{-1}$ \cite{2020A&A...641A...6P}. The second method involves directly measuring the expansion rate in the local Universe by observing the redshifts of SNe\,Ia and other distance indicators like Cepheid variable stars. These late-Universe measurements consistently yield a higher value of $H_0=73.04\pm1.04\rm\,km\,s^{-1}\,Mpc^{-1}$ \cite{2022ApJ...934L...7R}. This statistically significant mismatch, known as the Hubble tension, suggests potential gaps in our understanding of cosmology. Several proposed solutions to this tension include exploring dynamical dark energy models, utilizing new cosmic probes like GWs, improving the calibration of distance indicators using gravitational lensing, and other approaches (see comprehensive discussions in \cite{2021CQGra..38o3001D} and \cite{2023ARNPS..73..153K}). Ultimately, resolving the Hubble tension will likely require a combination of more precise observations and potentially a revision of the current cosmological framework.

To address this challenge, we consider FRBs as a potential tool for estimating $H_0$. There are  relatively recent studies conducted previously in this direction as mentioned in the Introduction. However, due to their consideration of fewer FRBs, they all have large error bars in their estimations of $H_0$. These error bars can, in principle, take into account for both $H_0$ obtained for the early- and late-time Universe cosmology, ultimately preventing to draw any decisive conclusion on the Hubble tension. In our study, we consider a larger sample of 64 extragalactic, localized FRBs from various telescopes. Employing Bayesian analysis with three different likelihood functions, we consistently obtain $H_0$ values well above $70\rm\,km\,s^{-1}\,Mpc^{-1}$, aligning with the late-Universe $H_0$ values. This is expected as the FRBs are localized to a maximum redshift of around 1, consistent with the typical redshifts considered in late-time Universe studies. It is noteworthy our estimated $H_0$ values have significantly reduced error bars of approximately $\pm 2.5\rm\,km\,s^{-1}\,Mpc^{-1}$, clearly distinguishing them from the early-Universe $H_0$, unlike previous FRB studies. Although the error bars may not be as tight as those derived from SNe\,Ia data, they are considerably smaller than those obtained from GW astronomy, demonstrating the potential of FRBs as a promising cosmological probe. 

\begin{figure}[htpb]
    \centering
    \includegraphics[scale=0.5]{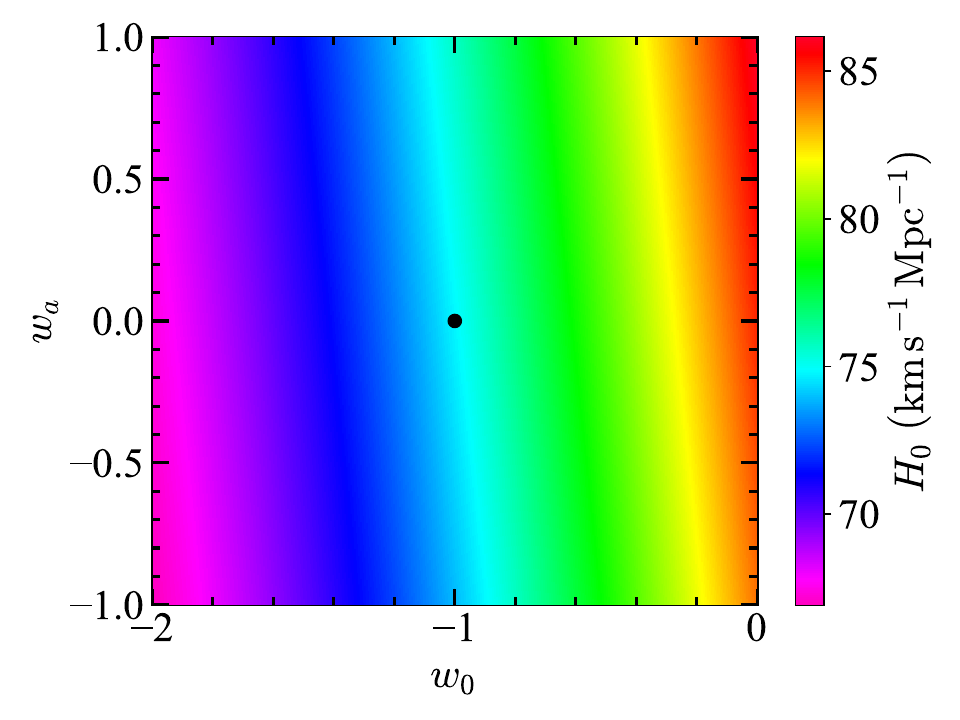}
    \caption{Variation of $H_0$ for different dark energy equation of states considering Model I for host galaxy DM distribution. The colorbar shows $H_0$ for different combinations of $w_0$ and $w_a$. The black point at the center corresponds to the Hubble constant value in $\Lambda$CDM cosmology.}
    \label{Fig: H0_MG}
\end{figure}
An important point to highlight is that the value of $H_0$ derived from FRB data is dependent on the underlying cosmological model. In our study, we have employed the $\Lambda$CDM framework. However, in the literature, various cosmological models have been considered, particularly in efforts to address the Hubble tension. One simple model involves a dynamical dark energy equation of state, defined as $w\equiv P/\rho$ where $P$ and $\rho$ represent pressure and energy density, respectively. A dynamical dark energy model has the potential to alleviate both the Hubble tension and the $\sigma_8$ tension simultaneously \cite{2020PhRvD.101l3516A,2021CQGra..38o3001D,2023PDU....4201266D}. For instance, we assume that the dark energy equation of state evolves over time, such that the time-dependent Hubble parameter can be expressed as \cite{2005PhRvD..72j3503J}
\begin{equation}
    H(z) = H_0\sqrt{\Omega_\mathrm{m} \left(1+z\right)^3 + \Omega_\Lambda f(z)},
\end{equation}
where $f(z) = (1+z)^{3\left(1+w_0+w_a\right)} \exp{-3 w_a z/(1+z)}$ with $w_0$ and $w_a$ being dimensionless parameters, and 
\begin{align}
    w(z) = w_0 + w_a \frac{z}{1+z}.
\end{align}
It is clear that for $w_0=-1$ and $w_a=0$ the standard $\Lambda$CDM cosmology with $w=-1$ is recovered. However, other parameter combinations can lead to $w(z)<-1$ for certain redshifts, which influences $\langle\mathrm{DM}_\mathrm{IGM}(z)\rangle$ and subsequently affects the likelihood functions. Figure~\ref{Fig: H0_MG} shows the estimated $H_0$ using the joint likelihood function of Equation~\eqref{Eq: Likelihood2}, assuming Model I for host DM distribution and different combinations of $w_0$ and $w_a$. We observe that, for certain parameter combinations, the alteration to the cosmological model leading to $w(z)<-1$ gives $H_0$ estimates closer to the early-Universe cosmology predicted values. This is in accordance with multiple recent cosmological surveys, like Planck, DESI, and Euclid, which suggest a deviation to $\Lambda$CDM model by fitting supernova, BAO, CMB, and weak lensing data \cite{2015PhRvD..91h3005H,2021A&A...654A.148M,2024JCAP...12..007C}. Note that different combinations of $w_0$ and $w_a$ can yield the same $H_0$. Hence the FRB data alone cannot put constraint on these parameters together with $H_0$ and it requires data from other observations.

\section{Conclusions}\label{Sec5}

In this study, we have explored the potential of FRBs as a tool for constraining the Hubble constant. Despite uncertainties in $\mathrm{DM}_{\mathrm{IGM}}$ and $\mathrm{DM}_{\mathrm{Host}}$ quantities, we have demonstrated that FRBs can serve as a valuable cosmological probe. By analyzing a dataset of 64 extragalactic, localized FRBs, we have consistently derived $H_0$ values exceeding $70\rm\,km\,s^{-1}\,Mpc^{-1}$ under $\Lambda$CDM cosmology, in agreement with the late-Universe $H_0$ estimates. Our results exhibit significantly reduced error bars, approximately $\pm 2.5\rm\,km\,s^{-1}\,Mpc^{-1}$, further distinguishing them from early-Universe measurements. Once the cosmology is deviated from the standard $\Lambda$CDM case, it is expected to have a different $H_0$ estimate. We have explored the dynamical dark energy model incorporating $w_0$ and $w_a$ parameters and showed that the result can match the early-Universe measurements for different combinations of these parameters. These findings emphasize the potential of FRBs to contribute to cosmological parameter constraints and highlight the ongoing Hubble tension.

Numerous cosmological probes and methods are currently under investigation to resolve the tensions in cosmology and enhance the precision of cosmological measurements. Gamma-Ray Bursts, Quasars, and Active Galactic Nuclei are being explored as potential standardizable candles to enable more accurate distance measurements \cite{2022A&A...663L...7S,2024Galax..12...69P}. Alternatively, BAOs are also being explored to use as standard ruler \cite{2016MNRAS.456L..45R}. Furthermore, novel approaches such as machine learning techniques and model-independent methods are being utilized to analyze large datasets and mitigate systematic biases. Independent methodologies, including Cosmic Chronometers and GW standard sirens, are also providing alternative measurements of the Universe's expansion rate. While alternative observational approaches, such as GW standard sirens from compact binary mergers and pulsar timing arrays, have been utilized to estimate $H_0$, their current limitations in source localization result in large uncertainties, limiting their ability to meaningfully address the tension \cite{2021ApJ...909..218A,2022MNRAS.517.1242M}. In contrast, our analysis demonstrates that FRBs, especially with precise localization and better statistical methodologies, can yield significantly tighter constraints on $H_0$. Given the rapidly increasing number of observed FRBs, including those with precise localization, this class of transients is emerging as a promising and complementary tool for cosmological studies. These further improve our understanding of key cosmological parameters and help resolve the $H_0$ discrepancy.

Upcoming surveys, such as Canadian Hydrogen Observatory and Radio-transient Detector (CHORD)\footnote{\url{https://www.chord-observatory.ca/}}, Hydrogen Intensity and Real-time Analysis eXperiment (HIRAX)\footnote{\url{https://hirax.ukzn.ac.za/}}, Square Kilometre Array (SKA)\footnote{\url{https://www.skao.int/en}}, Deep Synoptic Array (DSA)-2000\footnote{\url{https://www.deepsynoptic.org/overview}}, and Bustling Universe Radio Survey Telescope in Taiwan (BURSTT)\footnote{\url{https://www.burstt.org/}} are expected to detect a larger number of FRBs, with a significant fraction localized to their host galaxies. These surveys promise to deliver more stringent constraints on cosmological parameters, potentially offering critical insights into the Hubble tension. In conclusion, our study underscores the potential of FRBs as a powerful tool for probing the large-scale structure of the Universe. As research in this field continues to advance, we anticipate that FRBs will play an increasingly prominent role in refining our understanding of the cosmos, from the properties of IGM to the determination of fundamental cosmological parameters.


\section*{Acknowledgments}

The authors would like to thank the anonymous reviewer for their constructive comments to improve the manuscript content. S.K. acknowledges S. Hagstotz (Ludwig Maximilian University of Munich) for insightful discussions and for providing the code used to generate the mock catalog of FRBs. We gratefully acknowledge support from the University of Cape Town Vice Chancellor’s Future Leaders 2030 Awards programme which has generously funded this research and support from the South African Research Chairs Initiative of the Department of Science and Technology and the National Research Foundation. S.K. is funded by the National Science Centre, Poland (grant no. 2023/49/B/ST9/02777). A.W. acknowledges support from the ICTP through the Associates Programme and from the Simons Foundation (grant no. 284558FY19). Computations were performed using facilities provided by the University of Cape Town’s ICTS High Performance Computing team: \href{https://ucthpc.uct.ac.za/}{hpc.uct.ac.za} \cite{university_of_cape_town_2023_10021613}.

\bibliographystyle{elsarticle-num-names} 
\bibliography{Bibliography}

\end{document}